\documentstyle[12pt,epsf]{article}

\begin{document}
\begin{titlepage}
   %May 16  2002

   \begin{center}
   {\Large \bf  CP violating phase $\gamma$ and
   the partial widths asymmetry in  $B^- \to \pi^+ \pi^- K^-$
   and  $B^- \to K^+ K^- K^-$    decays  }\\
   \vspace{1cm}
   {\large \bf S. Fajfer$^{a,b}$, R. J. Oakes$^{c}$ and
   T.N. Pham$^{d}$\\}

   {\it a) J. Stefan Institute, Jamova 39, P. O. Box 3000, 1001 Ljubljana,
   Slovenia}
   \vspace{.5cm}

   {\it b)
   Department of Physics, University of Ljubljana, Jadranska 19, 1000
   Ljubljana,
   Slovenia}
   \vspace{.5cm}

   {\it c) Department of Physics and Astronomy, Northwestern University,
   Evanston, Il 60208, U.S.A.\\}
   \vspace{.5cm}

   {\it d) Centre de Physique Theorique, Centre National de la Recherche
   Scientifique,
   UMR 7644, Ecole Polytechnique, 91128 Palaiseau Cedex, France\\}

   \end{center}

   \vspace{0.25cm}

   \centerline{\large \bf ABSTRACT}

   \vspace{0.2cm}

   Motivated by recent Belle measurement of the
   branching ratios $B^- \to \pi^+ \pi^- K^-$
   and $B^- \to K^+ K^- K^-$ and $B^\pm \to \chi_{c0} K^\pm$,
   we investigate the CP violating  asymmetry in the partial widths
     for the decays  $B^- \to \pi^+ \pi^- K^-$
   and $B^- \to K^+ K^- K^-$, which results from the interference of the
   nonresonant decay amplitude with the resonant amplitude
   $B^\pm \to \chi_{c0} K^\pm $ $ \to\pi^+ \pi^- K^\pm  $ and
     $B^\pm \to \chi_{c0} K^\pm$ $ \to K^+ K^- K^-$.
     By taking  $\gamma \simeq 58^o$ we predict that the 
     partial widths asymmetry  is   reaching  $10 \%$ for the $B^- \to \pi^+ \pi^- K^-$ decay
     and $16\%$ for the  $B^- \to K^+ K^- K^-$decay.

   \end{titlepage}
   The extraction of the CP violation phases $\alpha$, $\beta$
   and $\gamma$ has stimulated many studies during last decade
   ((see e.g. \cite{neubert}).
   Among them the phase $\gamma$ is most difficult to constrain.
   Numerous proposals were aimed to determine  its size (see e.g.
   \cite{fleischer}).
   The partial width asymmetry in charmless three-body decays
   offers a chance to get more information on 
   $\sin \gamma$ \cite{EGM} - \cite{DGLNP}. 

   The basic idea was that the nonzero asymmetry results from the
   interference of the nonresonant three-body decay amplitude and
   the resonant
   $B^- \to \chi_{c0} \pi^- \to \pi^+ \pi^- \pi^-$ decay amplitude
   \cite{EGM,BFOPP,DEHT,DGLNP}.
   All these approaches were based on a theoretical prediction
   \cite{EGM} for the decay rate of  $B^- \to \chi_{c0} \pi^-$
   which is difficult to obtain in the factorization model.
   Instead of relying on this prediction for this decay rate, we can now
   use the
   recent  Belle collaboration measurement of
   the decay rate
   \cite{Belle1,Belle2}
   \begin{eqnarray}
   {\rm BR}(B^- \to \chi_{c0} K^-) = (6.0^{+2.1}_{-1.8})\times 10^{-4}.
   \label{b1}
   \end{eqnarray}
   which is surprisingly large, comparable to the $B^- \to J/\psi K^-$ decay
   rate.
   In addition, the Belle collaboration
   has anounced the measurement of the decay rates
   \begin{eqnarray}
     {\rm BR}(B^- \to K^+ K^- K^-) = (37.0\pm 3.9 \pm 4.4) \times 10^{-6},
   \label{b2}
   \end{eqnarray}
   and
   \begin{eqnarray}
     {\rm BR}(B^- \to \pi^+ \pi^- K^-) =
     (58.5\pm 7.1 \pm 8.8) \times 10^{-6}.
   \label{b3}
   \end{eqnarray}
   Although these measurements do not distinguish yet the contributions to
   the branching ratios of the nonresonant decays,
   they constrain their magnitudes. On the other hand,
   the charmless two-body decays of B mesons
   are stimulating many research projects.
   The factorization approximation is usually  used and recently
   great improvement has been made in the understanding
   of $B \to \pi \pi$ and $B \to \pi K$ decay modes
   (e.g. \cite{BBNS}).
   The decay mechanism of the nonleptonic charmless
   three-body B decays is much less understoood. The intermediate
   resonance states might be simpler to understand due to better
   knowledge of the two-body B decays.
   Due to the lack of understanding of the three-body
   nonresonant decay modes of B mesons,
   we will adopt the existing approaches \cite{BFOPP,DEHT}.
   The first is the use of the
   factorization approach. Then we reduce the problem to the
   calculation of the matrix elements of currents
   (or density operators). In the evaluation of these matrix elements,
   the main contribution to the nonresonant
   $B^{\pm} \to M {\bar M} K^{\pm}$, $ M = K^+, \pi^+$,
   amplitude comes from  the product
   $< M \bar M| ({\bar u} b)_{V - A}| B^->$ $ < K^- | ({\bar s} u)_{V -A} |0>$
   where $(\bar q_1 q_2 )_{V-A}$ denotes
   $\bar q_1 \gamma_{\mu} (1- \gamma_5)q_2$,
   or from
   $\sum_q< M \bar M| ({\bar q}(1 - \gamma_5) b| B^->$
   $ < K^- | ({\bar s}(1 + \gamma_5) q) |0>$.
   For the calculation of the matrix element
   $< M \bar M| ({\bar u} b)_{V - A}| B^->$
   we extend  the results obtained in \cite{BFOP},
   where the nonresonant
   $D^+ \to K^- \pi^+ l \nu$ decay
   was analyzed.
     In this analysis the experimental result for
   the branching ratio of
   the nonresonant $D^+ \to K^- \pi^+ l \nu$ decay was successfully reproduced
   within a hybrid framework  which combines the
   heavy quark effective theory (HQET)
   and the chiral Lagrangian (CHPT) approach.

   The combination  of heavy quark symmetry and  chiral symmetry has been
   quite successful in the
   analysis of D meson semileptonic decays \cite{BFOP} - \cite{BFO}.
   Heavy quark symmetry is expected to be even  better
   for the heavier B mesons \cite{caspr,wise}. However, CHPT
   could be worse in B decays due to the large
   energies of light mesons in
   the final state \cite{FOP}.
   In order to adjust the model to be applicable
   in the whole kinematic region  we  retain
   the usual HQET Feynman rules for the
   {\it vertices} near and outside the zero-recoil region, as in
   \cite{caspr,wise},
   {\it but we include the complete
   propagators instead of using the usual HQET propagator}
   \cite{BFOP,BFO,BFOS}.

   In the following we systematically use this model \cite{BFOP,BFO}
   to calculate the nonresonant
   $B^{\pm} \to M {\bar M} K^{\pm}$, $ M = K^+, \pi^+$, decay amplitude.
   We find that contrary to the previous cases \cite{FOP}  penguin
   contributions are
   very important.

   First we discuss the
   contributions to the nonresonant decay amlitude and then
   we discuss the partial width asymmetry.

   The effective weak Hamiltonian  for the nonleptonic Cabibbo-suppressed
   $B$ meson decays is given by 
   \cite{Ali} - \cite{BDHP}
   \begin{eqnarray}
   {\cal H}_{eff}&  = & \frac{G_F}{{\sqrt 2}} [V_{us}^* V_{ub}
   (c_1 O_{1u} +
   c_2 O_{2u} )  + V_{cs}^* V_{cb} (c_1 O_{1c} +c_2 O_{2c} ) ]\nonumber\\
   & - &\sum_{i=3}^{10} ([V_{ub} V_{ud}^* c_i^u
     +  V_{cb} V_{cs}^* c_i^c + V_{tb} V_{ts}^* c_i^t) O_i ] + h.c.
   \label{eq1}
   \end{eqnarray}
   where the superscripts $u$, $c$, $t$ denote the internal quark.
   The tree-level operators are defined as
   \begin{equation}
   O_{1q} = \bar q \gamma_{\mu} (1 - \gamma_5) b
   \bar s \gamma^{\mu} (1 - \gamma_5) q ,
   \label{O1}
   \end{equation}
   with $q= u,c$ and
   \begin{equation}
   O_{2q} = \bar q \gamma_{\mu} (1 - \gamma_5) q
   \bar s \gamma^{\mu} (1 - \gamma_5) b.
   \label{O2}
   \end{equation}
   We rewrite
   $O_3$ $-$ $ O_6$, using the Fierz transformations, as follows:
   \begin{equation}
   O_3 = \sum_{q = u,d,s,c,b} \bar s \gamma_{\mu} (1 - \gamma_5) b
   \bar q \gamma^{\mu} (1 - \gamma_5) q ,
   \label{O3}
   \end{equation}
   \begin{equation}
   O_4 = \sum_{q = u,d,s,c,b} \bar s \gamma_{\mu} (1 - \gamma_5) q
   \bar q \gamma^{\mu} (1 - \gamma_5) b,
   \label{O4}
   \end{equation}
   \begin{equation}
   O_5 = \sum_{q = u,d,s,c,b} \bar s \gamma_{\mu} (1 - \gamma_5) b
   \bar q \gamma^{\mu} (1 +  \gamma_5) q ,
   \label{O5}
   \end{equation}
   \begin{equation}
   O_6 = - 2 \sum_{q = u,d,s,c,b} \bar s (1 + \gamma_5) q
   \bar q  (1 -\gamma_5) b,
   \label{O6}
   \end{equation}
   The operators $O_7$,...,$O_{10}$ denote the electroweak penguin
   operators.
   Due to the smallness of their Wilson coefficients we do not
   include them in our analysis \cite{FOP,Ali1}.

   The factorization approximation is obtained by neglecting in the
   Lagrangian terms which are the product of two
   color-octet operators after Fierz reordering of the quark fields. The
     effective Lagrangian for non-leptonic decays are then given by
   (\ref{eq1}) with $c_{i}$ replaced by $a_{i}$. 
   In our calculations we use next-to-leading Wilson coefficents
   (e.g. \cite{Ali1}   \cite{IP}). In \cite{Ali1,Ali2,DES} 
   it was found that  $a_3$ and $a_5$ are one order of
   magnitude smaller  than $a_4$ and $a_6$
   and therefore we can safely neglect the contributions from $O_{3}$ and
   $O_{5}$ operators.
   For $N_{c}=3$, $m_b = 5 \rm \, GeV$, we use  \cite{DES,IP}~:
   \begin{equation}
   \begin{array}{rlrl}
   a_1=& \ 1.05,  &a_2= & \ 0.07, \\
   a_4= & -0.043- 0.016 i, \\
   a_6= & -0.054- 0.016 i.\\
   \end{array}
   \label{ai1}
   \end{equation}
  It is important to notice that the imaginary parts of the 
  Wilson coefficients for the penguin 
  operators are due to the internal charm- and up-quark loop
  exchanges. They introduce additional strong phases, affecting 
  the CP asymmetries.  
   
   The use of factorization gives for the matrix element of
   $O_1$ operator for the nonresonant $B ^- \to \pi^+ \pi^- K^-$
   decay, following the results of \cite{BFOPP}
   \begin{eqnarray}
   < \pi^+ (p_1)\pi^-(p_2) K(p_3)| O_1 | B^- (p_B)>_{nr}&= & \nonumber\\
   < K^- (p_3)|( {\bar s} u)_{V-A}| 0>
   < \pi^- (p_2) \pi^+ (p_1) | ({\bar u} b)_{V-A} | B^- (p_B)>_{nr} .
   \label{eFig1c}
   \end{eqnarray}
   To evaluate  the
   matrix element $< \pi^-(p_1) \pi^+(p_2) | ({\bar u} b)_{V -A}
   | B^- (p_B)>_{nr}$
   we will also use the results obtained previously
   in the analysis
   of the nonresonant
   $D^+ \to \pi^+ K ^- l \nu_l$ decay width \cite{BFOP}.
   We write the matrix element \break
   $< \pi^-(p_1) \pi^+(p_2) | ({\bar u} b)_{V -A}| B^- (p_B)>$
   in the general form
   \begin{eqnarray}
   < \pi^-(p_1) \pi^+(p_2) | ({\bar u} b)_{V -A}
   | B^- (p_B)>
   &\!\!\! = \!\!\!&
   ir(p_B-p_2-p_1)_\mu\nonumber\\
   +iw_+(p_2+p_1)_\mu+iw_-(p_2-p_1)_\mu
   &\!\!\! -\!\!\!&2h\epsilon_{\mu\alpha\beta\gamma}p_B^\alpha p_2^\beta
   p_1^\gamma\;.
   \end{eqnarray}
   The form factors $w_\pm^{nr}$  and $r^{nr}$ for the
   nonresonant decay are given in
   \cite{BFOPP,FOP,BFOP} together with a detailed description of
   our hybrid model.
   Using the following
   set of Mandelstam's variables: $s= (p_B - p_3)^2$,
   $t= (p_B - p_1)^2$ and $u= (p_B-p_2)^2$
   the formfactors can be written as
   \begin{eqnarray}
   w_+^{nr}(s,t) & = & - \frac{g}{f_1 f_2}
   \frac{f_{B*} m_{B*}^{3/2} m_B^{1/2}}{(t - m_{B*}^2} [ 1 -
   \frac{(m_B^2 -m_1^2 -t)}{ 2 m_{B*}^2} ]\nonumber\\
   & + & \frac{f_B}{ 2 f_1f_2}  -
   \frac{{\sqrt m_B} \alpha_2}{ f_1 f_2}
   \frac{1}{2 m_B^2}(2 t + s - m_B^2 -m_3^2 -2 m_1^2),
   \label{w+1-h}
   \end{eqnarray}
   \begin{eqnarray}
   w_-^{nr}(s,t) & = &  \frac{g}{f_1f_2}
   \frac{f_{B*} m_{B*}^{3/2} m_B^{1/2}}{ t  - m_{B*}^2}
   [ 1 + \frac{(m_B^2 -m_1^2 -t)}{2 m_{B*}^2}]\nonumber\\
   &  + &\frac{{\sqrt m_B} \alpha_1}{f_1f_2}.
   \label{w-1h}
   \end{eqnarray}

   \begin{eqnarray}
   r^{nr}(s,t) & = & - \frac{ 1 + \tilde \beta}{f_1 f_2}
   \frac{1}{2} (2 t + s - m_B^2 -m_3^2 -2 m_1^2)
   {\sqrt \frac{m_{B'}}{m_B}} \frac{f_{B'}}{
   m_3^2 - m_{B'}^2}
   \nonumber\\
   & - & {\sqrt \frac{m_B}{m_{B"}}}
   \frac{ 4 g^2 f_{B''} m_{B'}^* m_{B''}}{f_1 f_2}
   \frac{1}{m_3^2 - m_{B''}^2}
     \nonumber\\
   & \times & \frac{[\frac{1}{2}(s - m_1^2 - m_2^2)  -
   \frac{1}{4 m_{B*'}^2}
   (t + m_2^2 - m_3^2)
   (m_B^2 - m_1^2 -t)]}{t - m_{B'*}^2} \nonumber\\
   & + & \frac{ g}{f_1 f_2} {\sqrt \frac{m_B}{m_{B'}}}
   \frac{f_{B'^*}}{t - m_{B'*}^2} (m_B^2 -m_1^2 -t)
     \nonumber\\
   & + & \frac{f_B}{2f_1 f_2} + \frac{{\sqrt m_B} \alpha_2}{ f_1 f_2}
   \frac{1}{2 m_B^2}(2 t + s - m_B^2 -m_3^2 -2 m_1^2)
   \label{rh}
   \end{eqnarray}
   With the use  of these results we write down
   \begin{eqnarray}
   < K^-(p_3) \pi^+(p_1) \pi^-(p_2)| O_1| B(p_B)>_{nr} = & & \nonumber\\
   -[f_3 m_3^2 r^{nr}  + \frac{1}{2} f_3(m_B^2 - m_3^2 -s)w_+^{nr} & &
   \nonumber\\
   \frac{1}{2}f_3 (s + 2 t - m_B^2 - 2 m_1^2 - m_3^2)w_-^{nr}]
   \label{o1mv}
   \end{eqnarray}
   The parameters $\alpha_{1,2}$ are
     defined in \cite{BFO},  $g$ is the $B^* B \pi$ strong coupling,
   discussed  in \cite{caspr,wise,BFO}.
   Here  $ B'$, $B^{'*}$, $B''$ denote the relevant  $B$ meson poles, and
   $f_{1,2}$
     denotes the pseudoscalar meson decay constants. The coupling
   $\tilde \beta $ has been analyzed in \cite{BFOS} and found to be close
   to zero
   and therefore will be neglected.

   The operator $O_4$ has the same kind of decomposition as $O_1$,
   while
   the operator $O_6$ can be rewritten as  the product of
   density operators. The evaluation of the matrix elements like  $<  K| \bar q (1
   +\gamma_5) b |B>$ and 
   $< \pi \pi| \bar q (1 - \gamma_5) b |B>$ can then be reduced to
   the evaluation of the matrix elements  of the weak currents
   $< \pi \pi  | \bar q \gamma_{\mu} \gamma_5 b |B>$ and  
   $< K| \bar q \gamma_{\mu}  b|B>$ following the procedure 
   described in detail in \cite{FOP}.
   For the
   $\bar s (1 + \gamma_5) q$ scalar and pseudoscalar quark  density operator
   we use the CHPT result \cite{GL} and obtain
     \begin{equation}
   \bar s (1 + \gamma_5) q = - \frac{f_{\pi}^2}{2} {\cal B} U_{qs}^\dag
   \label{den}
   \end{equation}
   where ${\cal B}$ is a
   real constant
   expressed in terms of quark
   and meson masses; e.g., to lowest order
   $m_{K^0}^2 = {\cal B} (m_s + m_d)$ and
   $U = exp (i2 \Pi /f)$ where $ \Pi$ is a pseudoscalar
   meson matrix \cite{FOP}. We use the value $m_s= 150 $
   ${\rm MeV}$ for the s quark mass
   at the scale  $m_B$ mass \cite{FOP}, giving  ${\cal B} = 1.6 $
   ${\rm GeV}$.
   For the calculation of the density operator
   $\bar q  (1 -\gamma_5) b$   we use
   the relations \cite{Ali2}
   \begin{equation}
   \bar q \gamma_5 b = \frac{-i}{m_b}\partial_{\alpha}(\bar q \gamma^{\alpha}
   \gamma_5 b),
   \label{hsr}
   \end{equation}
   and
   \begin{equation}
   \bar q b = \frac{i}{m_b}\partial_{\alpha} (\bar q \gamma^{\alpha}  b),
     \label{hsrv}
   \end{equation}
   where $m_q$ has been dropped since $m_q << m_b$.

   The factorization assumption in the case of the $O_6$ operator
   results in:
   \begin{eqnarray}
   <K^-(p_3) \pi^+(p_1) \pi^-(p_2)| O_6| B(p_B)> & = & \nonumber\\
     -2  \sum_{u,d,s,c,b}
   < K |\bar s (1 +\gamma_5) u |0 >
     <\pi \pi| \bar u (1 - \gamma_5) b |B> .&& %\nonumber\\
    \label{o6fac}
   \end{eqnarray}
   Since the density operator in 
   $ < \pi K|\bar s (1+ \gamma_5) d |0 >$ 
   can be related to the matrix element of the corresponding 
   current, and  the form factors describing the matrix element
    of the current operator are dominated  by  light 
  vector meson resonances, contributions from terms of the type 
   $ < \pi K|\bar s (1+ \gamma_5) d |0 >$ 
   $< \pi | \bar d (1 - \gamma_5) b |B>$ 
  is suppressed in the high energy region for the nonresonant 
  amplitude, and is neglected here for our purpose. 
  There is also the contribution  
  $< K \pi \pi|\bar s (1 + \gamma_5) u |0 >$ 
  $< 0 | \bar u (1 - \gamma_5) b |B>$ and  
  the pole contribution of the type
  $< K \pi \pi|{\cal L}_s| K(p_B)> $ 
  $(m_B^2- m_K^2)^{-1}$ $<K| \bar s (1 + \gamma_5) u |0 >$ 
   $< 0 | \bar u (1 - \gamma_5) b |B>$. Here ${\cal L}_s$ is the chiral
   Lagrangian for the light pseudoscalar mesons. 
  These two terms give negligible contribution to the branching ratio as expected 
  from annihilation graphs in \cite{Ali}.
  The contribution $< K \pi \pi|\bar s (1- \gamma_5) b |B^- >$
  $<0| \bar s (1 + \gamma_5) s |0 >$  
  does not contribute due to the momentum conservation.
  
  Using the $s$ and $t$ variables we derive:
   \begin{eqnarray}
   < K^-(p_3) \pi^+(p_1) \pi^-(p_2)| O_6| B(p_B)>_{nr} = & & \nonumber\\
   \frac{ {\cal B}}{m_B}\{ -2 \frac{f_1 f_2}{f_3} m_3^2 r^{nr}
   - \frac{f_1 f_2}{f_3} (m_B^2 - m_3^2 -s)w_+^{nr}& & \nonumber\\
   -  \frac{f_1 f_2}{f_3} (s + 2 t - m_B^2 - 2 m_1^2- m_3^2 )w_-^{nr}
   \}.
   \label{o6mv}
   \end{eqnarray}
   The dominant contributions to the
   amplitude for $B^- \to K^- \pi^+ \pi^-$ decays are
   then
   \begin{eqnarray}
   {\cal M}_{nr} = \frac{G}{{\sqrt 2}}[ V_{ub} V_{us}^* a_1
   <K \pi \pi | O_1| B>_{nr} - V_{tb} V_{ts}^* ( a_4
   <K \pi \pi | O_4| B>_{nr}
   + &&\nonumber\\
   a_6 <K \pi \pi | O_6| B>_{nr} )].
   \label{amp}
   \end{eqnarray}
   The decay width of the nonresonant
   $B^- \to \pi^- \pi^+ K^- $ can be found  with the help of
   \begin{equation}
   \Gamma_{nr} (B^- \to \pi^- \pi^+ K^-) = \frac{1}{(2 \pi)^3}
   \frac{1}{32 m_B^3} \int |{\cal M}_{nr}|^2~  ds~ dt.
   \label{dw}
   \end{equation}
   In the case of $B^- \to K^- K^+ K^- $ one has to take into account the two
   identical mesons
   in the final state (factor $1/2$  in the branching ratio)
     and to include in the amplitude the additional
     contribution obtained by
     the replacement $s \leftrightarrow t$.

   The lower and the upper bounds are given by
   $s_{min} = (m_1 + m_2)^2$, $ s_{max} = (m_B - m_3)^2$,  while for $t$
   they are given by
   \begin{eqnarray}
   t_{min,max} (s) & = & m_2^2 + m_3^2 - \frac{1}{s} [
   (s - m_B^2 + m_3^2)(s + m_2^2 - m_1^2)\nonumber\\
   & \mp &\lambda^{{1\over 2}}( s, m_B^2,m_3^2)
   \lambda^{{1\over 2}}(s, m_2^2,m_1^2)],
   \label{bounds}
   \end{eqnarray}
   where $\lambda(a,b,c) = a^2 + b^2 + c^2 -2( ab + ac + ab)$.

   The Wolfenstein parametrization of the CKM matrix \cite{wolf} gives
   $V_{ub} = A \lambda^3(\bar \rho - i \bar \eta)$,
   $V_{us}= \lambda $, $V_{cs} = 1 - \lambda^2/2$,
   $V_{cb} =  A \lambda^2$,
   $V_{ts}= - A\lambda^2 $ and $V_{tb}= 1$.
   The values $ A= 0.82$, $\lambda = 0.224$, are well known, while
   for  $\bar \rho $ and  $\bar \eta$  there are bounds given in
   \cite{beneke}.
   This parametrization indicates $V_{ub}$ contains the phase,
     while the
   combination of $V_{ts}^*V_{tb}$ has none.
   We denote $V_{ub}= |V_{ub}| e ^{i\gamma} $ following
   \cite{EGM}.
   In our numerical calculation  we use the recent value
   $f_B = 175$ $\rm{ MeV}$ (see e.g. \cite{sahrajda}).
   The strong coupling $g$ is determined by the CLEO experimental
   results on the $D^*$ decay width \cite{CLEO1}.
   The value $g=0.57$ reproduces this decay width.
   Heavy quark symmetry requires that this parameter be the
   same for B mesons.
   However, it has been pointed out that the off-
   mass-shell
   effects are very important in view of the higher B meson energy scale
   \cite{nang} and therefore we will use the lower value of
   $g =0.23$ \cite{BFOPP,FOP,stewart}. This smaller value of g correctly
   reproduces
   the form factor for the
   $B \to \pi$ transition at zero momentum transfer, calculated in this model,
   \cite{BFO} in agreement
   with results obtained in other approaches (e.g. \cite{Ali1}).
   We fix the parameters $\alpha_{1,2}$ to repropduce
   the value $|A_{1,2}^{DK*} (0)| $ in \cite{PDG}.
   These results are obtained by assuming a pole behaviour
   of form factors. We follow this procedure and
   by using the values of
   these form factors at $Q^2_{max}$, we obtain
   $\alpha_1^{DK*}=  0.16$$[{\rm GeV}^{1/2}]$ and
   $\alpha_2^{DK*}= 0.05$$[{\rm GeV}^{1/2}]$, which  after soft
   scaling described in
   \cite{BFOPP,caspr} give for the B mesons
   $\alpha_1^{DK*} = \alpha_1^{B\rho}$ and  $\alpha_2^{DK*}/m_D =
   \alpha_2^{B\rho} /m_B$ resulting in the values
   $\alpha_1= 0.16$ $ [{\rm GeV}^{1/2}]$,
   $\alpha_2= 0.15$ $[{\rm GeV}^{1/2}]$ which we will use here. These values
   give
   $A_{1,2}(0)$
   for the $B \to \rho$ transiton in agreement with the values
   used in \cite{Ali}.
   The branching ratio can be written as a sum of tree level contribution
   $T$ (in which the operator $O_1$ contributes), the 
   penguin contributions $P$ (operators $O_4$ and $O_6$) 
   and there are two
   interference terms proportional to
   $ \cos \gamma$  (denoted by  $I_1$) and $\sin \gamma$  
   (denoted by $I_2$): 
     \begin{equation}
   {\rm BR}(B^-\to K^- \pi^+ \pi^-)_{nr} = T + P + I_1 \cos \gamma
   + I_2 \sin \gamma. 
   \label{kpp-b1}
   \end{equation}
    After performing the  numerical integrations using 
    the parameters described above, 
   we obtain 
   $T = 7.0\times 10^{-6}$, $P= 7.5 \times 10^{-5}$, 
   $I_1 = -4.3 \times 10^{-5}$ and  $I_2 = -1.5 \times 10^{-5}$.

  In the case of $B^- \to K^- K^+ K^- $ decay we find for
    $T = 3.4\times 10^{-6}$, $P= 3.7\times 10^{-5}$, 
   $I_1 = -2.1 \times 10^{-5}$ and  $I_2 = -7.4 \times 10^{-6}$. 
   Taking a larger value $g\simeq 0.4$, as some of typical smaller
   value of $g$ as mentioned  in
    \cite{damir}, we find 
   moderate increase of all values $T$, $P$, $I_{1,2}$ by about factor $20 \%$. 
 Note that while the penguin contributions dominate
   in these decays, the size of $\gamma$ is  very important for
   the magnitude of the branching ratio.

     The results obtained by Belle collaboration  \cite{Belle2}
   are ${\rm BR}(B^-\to K^- \pi^+ \pi^-)_{exp} $ $ =
   (55.6 \pm 5.8\pm 7.7)\times 10^{-6}$ and
   ${\rm BR}(B^- \to K^- K^+ K^- )_{exp}$ $  =
   (35.3 \pm 3.7\pm 4.3)\times 10^{-6}$.
   Comparing our results with these data we conclude that
   larger values of $g$ (e.g. $g\simeq 0.4$)
   should be discounted in these cases; they give rates for nonresonant
   decays larger than the
   experimental results.
   \\

Due to the strong phases in $a_4$ and $a_6$ 
one can generate the CP violating asymmetry as
 \begin{eqnarray}
A = \frac{| \Gamma (B^- \to  \pi^+ \pi^- K^-) - 
\Gamma (B^+ \to  \pi^+ \pi^- K^+)|}{ |\Gamma 
(B^- \to  \pi^+ \pi^- K^-) + 
\Gamma (B^+ \to  \pi^+ \pi^- K^+)|},  
\label{asint}
   \end{eqnarray}
which can be written in the form
\begin{eqnarray}
A = \frac{ \sin \gamma N_1}{N_2 + \cos \gamma N_3},  
\label{asint1}
   \end{eqnarray}
where 
\begin{eqnarray}
N_1 = - 2 G^2 |V_{us}^*||V_{ub}| |V_{ts}^*||V_{tb}|
\int (d PS)a_1 <O_1> [ {\rm Im} (a_4) <O_4> + && \nonumber\\
{\rm Im} (a_6) <O_6>], 
\label{N1}
\end{eqnarray}
\begin{eqnarray}
N_2 = G^2 |V_{us}^*|^2|V_{ub}|^2 \int (d PS)\{a_1| <O_1>|^2 
+ |V_{ts}^*|^2|V_{tb}|^2 [|a_4|^2 <O_4> +&& \nonumber\\ 
|a_6|^2 <O_6>]\}, 
\label{N2}
\end{eqnarray}
\begin{eqnarray}
N_3 = -2 G^2 |V_{us}^*||V_{ub}| |V_{ts}^*||V_{tb}|
\int (d PS)a_1 <O_1> [ {\rm Re} (a_4) <O_4> +&& \nonumber\\ {\rm Re} (a_6) <O_6>].  
\label{N3}
\end{eqnarray}   
  
 The numerical calculation
 gives for the $B^- \to  \pi^+ \pi^- K^-$ decay 
 $N_1= -3.0 \times 10^{-5}$, $N_2= 16,4 \times 10^{-5}$  
 and  $N_3= -8.6\times 10^{-5}$.
 For the $B^- \to  K^+ K^- K^-$decay  it becomes
$N_1= -1.5 \times 10^{-5}$, $N_2= 8.2 \times 10^{-5}$  
 and  $N_3= -4.2\times 10^{-5}$. 
 One might think  that the  nonresonant decay amplitude might 
 have  more 
 complicated structure than what we found within this model
 and then the CP violating asymmetry will be  changed.  
 However,  in  the  
 factorization model, which we use in the above calculation, the CP
 violating asymmetry in the nonresonant partial or integrated decay
 rates, is essentially model independent, since the matrix elements of
 $O_{4}$ and $O_{6}$ are practically proportional to that of $O_{1}$, as 
 can be seen from Eqs (\ref{o1mv}) and (\ref{o6mv}). Even in  
 the presence of low-energy 
 resonances contribution in the nonresonant amplitude, the CP asymmetry,
 for the total decay rates integrated over the whole phase
 space, or for the differential decay rates measured as a function 
 of the two-pion or two-kaon
 invariant mass, can be computed in terms of the Wilson coefficients
 $a_{4}$ and $a_{6}$, and are independent of the form factors. The CP 
 violating asymmetry Eq.(\ref{asint}) depends on the strong phase generated
 by the absorptive part of the Wilson coefficients. 
 
 On the other hand,
 the partial widths asymmetry at the charmonium resonance discussed 
in \cite{EGM,BFOPP, FOP}
 seems to be more reliable for a further constraint of the CP 
 violating phase $\gamma$. 
  
 In order to obtain the  partial width CP asymmetry, one also needs  to
   calculate the resonant decay amplitude
   $B^- \to $ $ \chi_{c0} K^-$
   $ \to \pi^+ \pi^- K^-$. This amplitude
   can  be easily determined  using the narrow width approximation,
   as in \cite{DEHT}:
   \begin{eqnarray}
   {\cal M}_{r}(B^- \to  \chi_{c0} K^- \to \pi^+ \pi^- K^-) & = &\nonumber\\
   {\cal M}(B^{-} \to \chi_{c0} K^- ) \frac{1}{ s - m_{\chi_{c0}}^2 +  i
   \Gamma_{\chi_{c0}} m_{\chi_{c0}}}
   {\cal M}( \chi_{c0} \to \pi^+ \pi^- ).
   \label{ares}
   \end{eqnarray}
   In our numerical calculations we will use the  recent
   Belle measurement
   of the branching ratio  \cite{Belle1}
   ${\rm BR}(B^{\pm} \to $ $\chi_{c0}$ $ K^{\pm} )$ $ =(6.0\pm1.1)\times
   10^{-4}$. The amplitude is then
   ${\cal M} (B^- \to \chi_{c0} K^- ) = 3.37 \times 10^{-7}$ ${\rm GeV}$.
   The $\chi_{c0}$ decay data \cite{PDG}
   then fix
   the decay amplitudes for
   $|{\cal M} (\chi_{c0} \to \pi^-\pi^+)| = 0.113$ ${\rm GeV}$ and
   $|{\cal M} (\chi_{c0} \to K^-K^+)| = 0.126$ ${\rm GeV}$.
     \vspace{0.5cm}

   The partial decay width $\Gamma_p$ for
   $B^- \to M {\bar M} K^- $, $M = \pi^+$, $K^+$,  which contains
   both the nonresonant and resonant
   contributions,  is obtained then by
   integration from $ s_{min} = (m_{\chi_{c0}} - 2\Gamma_{\chi_{c0} })^2$ to
   $ s_{max} = (m_{\chi_{c0}} + 2\Gamma_{\chi_{c0} })^2$, where
   $m_{\chi_{c0}} = 3.415~\, \rm GeV$ and  $\Gamma_{\chi_{c0} } = 0.0149
   ^{+0.0026}_{-0.0023}~\, \rm GeV$ is the width of the  $
   \chi_{c0}$:
   \begin{equation}
   \Gamma_p = 
   \frac{1}{(2 \pi)^3} \frac{1}{32 m_B^3}
   \int_{s_{min}}^{s_{max}} ds \int_{t_{min}(s)}^{t_{max}(s)} dt~|{\cal
   M}_{nr } +
   {\cal M}_{r} |^2 .
   \label{dwp}
   \end{equation}
   Similarly, $\Gamma_{\bar p}$, the partial decay width for
   $B^+ \to  M {\bar M} K^+$, $M = \pi^+$, $K^+$ also  contains both  the
   nonresonant and
   resonant contributions.
     The CP nonconserving asymmetry
   is defined  by
   \begin{equation}
   A_p = \frac{ |\Gamma_p - \Gamma_{\bar p}|}{|\Gamma_p +
   \Gamma_{\bar p}|}.
   \label{asym}
   \end{equation}
   In calculation of the $\Gamma_p - \Gamma_{\bar p}$
   we derive, asumming that $V_{ub} = |V_{ub}|e^{i \gamma}$ :
   \begin{eqnarray}
     &&\Gamma_p - \Gamma_{\bar p} =
   \sin \gamma \frac{4 m_{\chi_{c0}}\Gamma_{\chi_{c0}}}{(2 \pi)^3 32 m_B^3}
     \nonumber\\
     &&\times \int_{s_{min}}^{s_{max}} ds \int_{t_{min}(s)}^{t_{max}(s)} dt~
   \frac{G}{{\sqrt 2}}|V_{ub}| |V_{us}^*|a_1
   <K \pi \pi | O_1| B>_{nr}\nonumber\\
   &&\times|{\cal M} (B^- \to \chi_{c0} K^- )|
   \frac{1}{ (s- m_{\chi_{c0}}^2)^2 + (m_{\chi_{c0}} \Gamma_{\chi_{c0}})^2}
   |{\cal M} (\chi_{c0} \to \pi^- \pi^+)|.\nonumber\\
   \label{asym-}
   \end{eqnarray}
   The denominator is given by
   \begin{eqnarray}
   &&\Gamma_p +  \Gamma_{\bar p} =
   2 \frac{1}{(2 \pi)^3 32 m_B^3}
     \int_{s_{min}}^{s_{max}} ds \int_{t_{min}(s)}^{t_{max}(s)} dt~\nonumber\\
   &&\times\{ |{\cal M}_{nr}|^2+ |{\cal M} (B^- \to \chi_{c0} K^- )
   \frac{1}{ s- m_{\chi_{c0}}^2 +  i m_{\chi_{c0}} \Gamma_{\chi_{c0}}}
   {\cal M} (\chi_{c0} \to \pi^- \pi^+)|^2\}.\nonumber\\
   \label{asym+}
   \end{eqnarray}
   Then we  can write   
   \begin{eqnarray}
   A_p = \frac{ A_1 \sin \gamma}{ A_2 + A_3 \cos \gamma},
   \label{ap}
    \end{eqnarray}
  where $A_1$ is determined by (\ref{asym-}), while 
  $A_2$ contains the sum of the resonant decay amplitude as well as
  $\gamma$-independent part of (\ref{asym+}), corresponding to 
  $N_2$ with the integration over the $\chi_{c0}$  resonance 
  region,  $ s_{min} = (m_{\chi_{c0}} - 2\Gamma_{\chi_{c0} })^2$, 
   $ s_{max} = (m_{\chi_{c0}} + 2\Gamma_{\chi_{c0} })^2$). 
  The part which is $\gamma$ dependent  is given in 
  $A_3$ (corresponding to 
  $N_3$ with the integration over the $\chi_{c0}$  
  resonance region).   
  Note that the term proportional to $\cos \gamma$ arising from 
  the interference with the resonance in the 
  denominator does not
   contribute. Namely, in the integration region of $s$
   in (\ref{asym+}),   the real part of
   the resonance amplitude, being
   antisymmetric in $s$ gives no contribution to the integrated decay rates
   over the $\chi_{c0}$ resonance.
   
   In  Fig.1 and Fig.2 we present the distribution $d \Gamma_p/ d s$
   in the region\break  $ [s_{min}, s_{max}]$. It is clear from these figures that
   the nonresonant contribution is approximately constant over this region.
   For the given set parameters,  with $g = 0.23$,
   the CP violating  asymmetries are
   \begin{eqnarray}
   A_p(B^\pm\to K^\pm \pi^+ \pi^-) = \frac{7.9 \sin \gamma}{
   73 - 1.2 \cos \gamma}, 
   \label{a-pi}
   \end{eqnarray}
   and
   \begin{eqnarray}
   A_p(B^\pm\to K^\pm K^+ K-) = \frac{ 7.2 \sin \gamma}{ 
  41 - 5.6 \cos \gamma} 
   \label{a-K}
   \end{eqnarray}
Following recently given bound for 
the phase $\gamma =( 58 \pm 24)^o$ \cite{neubert} we find that 
the partial widths 
asymmetry for the $B^\pm\to K^\pm \pi^+ \pi^-$ decay is reaching $10\%$ 
and $ 16 \%$ for the $B^\pm\to K^\pm K^+ K-$. 

   The uncertainties due to the errors in
   the remaining input parameters have not been included here, but we can
   roughly estimate that
     the  error in the  asymmetry  can be as large as $40\%$.\\

   In summary, we can compare the previously considered cases
   $B ^\pm \to M \bar M  \pi^\pm$, $M = \pi^+, K^+$
   \cite{BFOPP,FOP} with the present cases
   $B^\pm\to  M \bar M K^\pm  $, $M = \pi^+, K^+$. Due to the
   different CKM matrix elements
   the penguin contributions
   dominate in the nonresonant 
   $B^\pm\to  M \bar M K^\pm  $, $M = \pi^+, K^+$ decay rate. Moreover,
   the phase $\gamma$ crucially influences  their magnitudes.
   The appearence of the strong phases in the $a_4$ and $a_6$ 
   Wilson coefficients 
   of the penguin operators can generate the CP violating 
   asymmetry in  the nonresonant decay modes  
   $B^\pm\to  M \bar M K^\pm  $, $M = \pi^+, K^+$. 
   
   The CP violating partial widths asymmetry is smaller in the cases
   $B^\pm\to  M \bar M K^\pm  $, $M = \pi^+, K^+$ than
   $B^\pm \to M \bar M  \pi^\pm$, $M = \pi^+, K^+$. 
   One might expect that, since now the CP violating  partial widths asymmetry
   comes from the interference
   of the resonant  amplitude and the tree-level nonresonant amplitude,
   which is given by $V_{ub}V_{us}^{*}$ which is CKM suppressed,
   relative to $V_{ub}V_{ud}^{*}$ for the decays
   $B^\pm \to M \bar M  \pi^\pm$, $M = \pi^+, K^+$. One could  also measure
   the differential decay rates and CP asymmetry as a function of the
   two-pion and two-kaon invariant mass squared as shown in Fig.1 and
   Fig.2 to obtain more informations on the CP asymmetry as well as
   possible determination of the nonresonant amplitudes at the 
   $\chi_{c0}$ mass region.
   
   The determination of these  partial width  asymmetries provides
   useful guidance for experimental searches 
   of CP violating effects and  further constraints on the
   phase  $\gamma$.\\

   This work was supported in part by the
   Ministry of Education, Science and Sport
   of the Republic of Slovenia (S.F), and by the U.S.
   Department of Energy, Division of High Energy Physics under grant
   No. DE-FG02-91-ER4086 (R.J.O.).
   S.F. thanks the   Centre de Physique Theorique,
   Ecole Polytechnique,
   for the warm hospitality during her stay there where
   part of this work was done.

   \newpage
   \vspace{-2cm}
   \begin{figure}[hbp]
   \centering
   \epsfxsize=6cm
   \epsffile{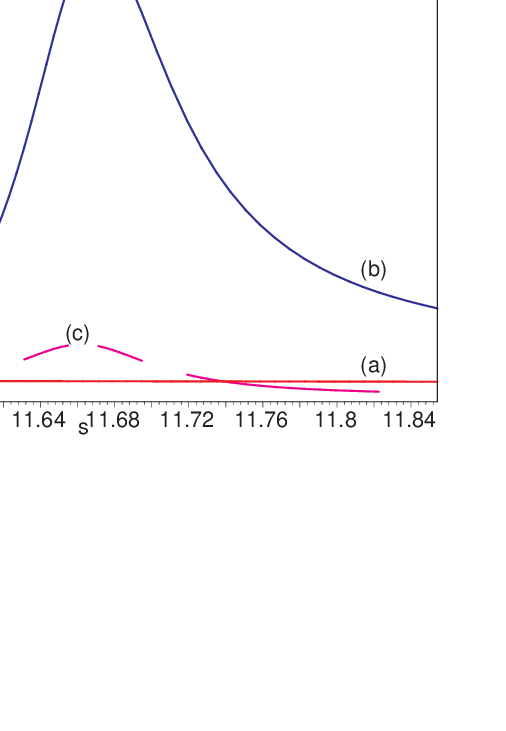}
   \vspace{-2cm}
   \caption{Differential branching ratios for
   $B^{-}\to \pi^{+}\pi^{-}K^{-}$ vs. $s$.
   The curves  (a), (b), (c) are
   $d({\rm B})(\rm NR)/ds$, $d({\rm B})/ds + d(\bar{{\rm B}})/ds$,
     $d({\rm B})/ds - d(\bar{{\rm B}})/ds$
   against the two-pion invariant mass squared s, for the nonresonant,
   CP symmetric and CP antisymmetric differential branching ratios
     respectively.}
   \label{Fig.1}
   \end{figure}
   \samepage
   \vspace{-2cm}
   \begin{figure}[hbp]
   \centering
   \epsfxsize=6cm
   \epsffile{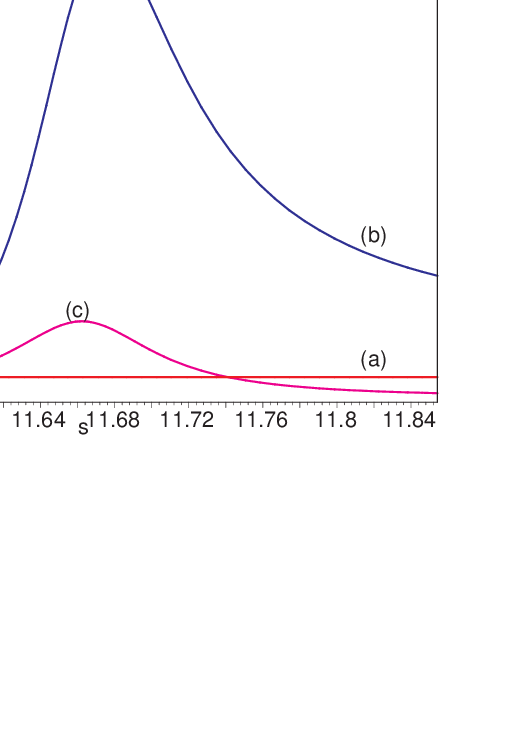}
   \vspace{-2cm}
   \caption{Differential branching ratios for
   $B^{-}\to K^{+}K^{-}K^{-}$ vs. $s$.
   The curves  (a), (b), (c) are
   $d({\rm B})(\rm NR)/ds$, $d({\rm B})/ds + d(\bar{{\rm B}})/ds$,
     $d({\rm B})/ds - d(\bar{{\rm B}})/ds$
   against the two-kaon invariant mass squared s, for the nonresonant,
   CP symmetric and CP antisymmetric differential branching ratios
     respectively.}
   \label{Fig.2}
   \end{figure}

   \end{document}